\documentclass{aa}

\usepackage{graphicx}
\usepackage{subfig}
\usepackage{epsfig}
\usepackage{epstopdf}
\usepackage{enumitem}
\usepackage{txfonts}
\usepackage{hyperref}
\usepackage{mathtools}

\def\lya{{Ly$\alpha$}}
\def\lyb{{Ly$\beta$}}
\def\keV{{\textrm{keV}}}
\begin{document} 

\title{Constraining the geometry of the nuclear wind in PDS 456 using a novel emission model}

\author{
A. Luminari\inst{1,2,3},
E. Piconcelli\inst{2},
F. Tombesi\inst{1,4,5},
L. Zappacosta\inst{2},
F. Fiore\inst{6},
L. Piro\inst{7},
F. Vagnetti\inst{1}
}

\institute{
Department of Physics, University of Rome ``Tor Vergata'', Via della Ricerca Scientifica 1, I-00133 Rome, Italy
\and
INAF Rome Astronomical Observatory, Via Frascati 33, 00078 Monteporzio Catone, Italy
\and
Department of Physics, University of Rome ``Sapienza'', Piazzale Aldo Moro 5, I-00185, Rome, Italy
\and
Department of Astronomy, University of Maryland, College Park, MD 20742, USA
\and
X-ray Astrophysics Laboratory, NASA/Goddard Space Flight Center, Greenbelt, MD 20771, USA
\and
INAF Trieste Astronomical Observatory, Via G. B. Tiepolo 11, I-34143 Trieste, Italy
\and
INAF Istituto di Astrofisica e Planetologia Spaziali, via Fosso del Cavaliere 100, I-00133 Rome, Italy
}

\date{Received June 12, 2018; accepted September 06, 2018}

\abstract
{Outflows from active galactic nuclei (AGN) are often invoked to explain the co-evolution of AGN and their host galaxies, and the scaling relations between the central black hole mass and the bulge velocity dispersion. Nuclear winds are often seen in the X-ray spectra through Fe K shell transitions and some of them are called ultra fast outflows (UFOs) due to their high velocities, up to some fractions of the speed of light. If they were able to transfer some percentage of the AGN luminosity to the host galaxy, this might be enough to trigger an efficient feedback mechanism.}
{We aim to establish new constraints on the covering fraction and on the kinematic properties of the UFO in the powerful ($L_{bol}\sim 10^{47}$ \textrm{erg/s}) quasar PDS 456, an established Rosetta stone for studying AGN feedback from disk winds. This will allow us to estimate the mass outflow rate and the energy transfer rate of the wind, which are key quantities to understand the potential impact on the host galaxy.}
{We analyze two sets of simultaneous XMM-Newton and NuSTAR observations taken in September 2013 and reported in Nardini et al. (2015) as having similar broadband spectral properties. We fit the Fe K features with a P-Cygni profile between 5 and 14 \textrm{keV}, using a novel Monte Carlo model for the WINd Emission (WINE).}
{We find an outflow velocity ranging from 0.17 to 0.28 \textrm{c}, with a mean value of 0.23 \textrm{c}. We obtain an opening angle of the wind of $71^{+13}_{-8}$ \textrm{deg} and a covering fraction of $0.7^{+0.2}_{-0.3}$, suggesting a wide-angle outflow. We check the reliability of the WINE model by performing extensive simulations of joint XMM-Newton and NuSTAR observations. Furthermore, we test the accuracy of the WINE model in recovering the geometrical properties of UFOs  by simulating observations with the forthcoming Advanced Telescope for High-Energy Astrophysics (ATHENA) in the X-ray band.}{}


\titlerunning{Constraining the geometry of the nuclear wind in PDS 456}
\authorrunning{A. Luminari et al.}
\maketitle

\section{Introduction}
Accretion disk winds are believed to play a fundamental role in the feedback from  active galactic nuclei (AGN) to their host galaxy. Both theoretical and observational evidences show that they are potentially able to transfer a significant fraction of the AGN's power (\citealp{di matteo}), up to $\sim 3-5\%\ L_{AGN}$ (\citealp{king_pounds15}, \citealp{tombesi15}, \citealp{Mrk231}, \citealp{WISSH I}, \citealp{ff17}). This amount of energy could overcome the binding energy of the host galaxy, and is often invoked to explain the observed relation between the supermassive black hole (SMBH) mass and the bulge velocity dispersion (the ``$M-\sigma$'' relation, see e.g., \citealp{kormendy}).
According to the most accepted scenario (see, e.g., \citealp{king_pounds15}), the gas is accelerated at accretion disk scales, propagates towards the host galaxy, and impacts the interstellar matter (ISM), producing a shock front (\citealp{zubovas}, \citealp{faucher}). If the kinetic energy is conserved during this process, the shocked gas may drive a massive, galaxy-scale outflow, with mass transfer rates up to $\sim 10^3$ \textrm{M$_{\odot}$ yr$^{-1}$} and velocities $=1000$ \textrm{km s$^{-1}$}.

In this two-phase scenario, the covering fraction of the disk wind, $C_f$, together with its velocity, density, and launching radius, are fundamental quantities to understand the amount of momentum and energy deposited into the ISM.
The accurate measurement of these quantities is crucial to reliably constrain the role of the disk winds in the co-evolution of AGN and their host galaxies.

In this paper, we present the application of our novel model of AGN wind emission (WINE) to the broadband XMM-Newton+NuSTAR spectrum of PDS 456. This nearby ($z=0.184$) luminous quasar exhibits the prototype of an ultra fast outflow (UFO) traced by a P-Cygni feature, due to highly-ionized Fe, with a ionization parameter $\xi=\frac {L_{ion}}{nR^2} \sim 10^5$ \textrm{erg\  cm\ s}$^{-1}$, where $L_{ion}$ is the luminosity in the 1-1000 Rydberg interval ($1 Rd=13.6$ \textrm{eV}), $n$ is the gas number density, and $R$ is the distance from the ionizing source.
The outflow velocity is $\sim 0.25$\textrm{c} and the launching radius is $< 200\  r_g$ \footnote{The gravitational radius $r_g$ is defined as $r_g=GM/c^2$, with \textrm{G} the gravitational constant and $M$ the black hole mass.} (\citealp{reeves03}, \citealp{nardini}, hereafter N15). 
N15 inferred a $C_f$ of the UFO of 0.8 (= 3.2$\pi$) and a kinetic power of $\sim 20\% $ of the bolometric luminosity of the quasar. These estimates are based on a spectral fitting using XSTAR tables \citep{kallman} to model both UFO emission and absorption features, in which the outflowing gas is approximated as a spherical shell with constant radial velocity.

The present paper is organized as follows. In Sect. 2 we describe the wind model. In Sect. 3 we present the X-ray data analysis and the results of the spectral fitting obtained by applying our model. Section 4 is devoted to the discussion of our findings, with particular emphasis on $C_f$ and the velocity of the outflow, the derived mass and energy transfer rates, and their implications for AGN feedback.

\section{The Wind Emission (WINE) model}
\label{wind model}
\label{Emission model}

\begin{figure}
\centering
\includegraphics[width=8cm]{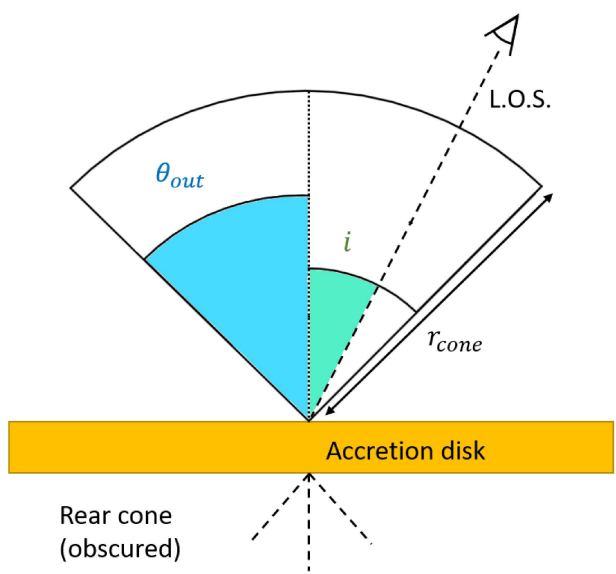}
\caption{Sketch of the WINE model. Further details can be found in Sect. \ref{wind model}.}
\label{Figmodelsketch}
\end{figure}

In the WINE model the wind is approximated as a conical region, with the vertex centered on the SMBH, the same symmetry axis as the accretion disk, and the gas velocity directed radially outward. This conical shape is consistent with the most popular accretion disk wind simulations (e.g., \citealp{proga}, \citealp{fukumura10}, \citealp{oshuga09}). As the UFOs originate from the innermost region of the accretion disk, we consider that the rear cone is not observable (see Fig. \ref{Figmodelsketch}).

The free parameters of the model are the opening angle of the cone (i.e., $\theta_{out}$ in Fig. \ref{Figmodelsketch}), the inclination angle $i$ of the line of sight (LOS) with respect to the symmetry axis, the maximum velocity of the wind ($v_{max}$), and the deceleration factor of the wind radial velocity ($s$). The radial velocity of the wind is defined as 
\begin{equation}
v(r') = v_{max} (1 - s r')
\label{eq_velocity}
,\end{equation}
where $r' \equiv \frac {r}{r_{cone}}$ and $r_{cone}$ is the height of the cone. Equation \ref{eq_velocity} represents a first-order expansion of the velocity with respect to $r$ and provides a basic description of the wind kinematic.\footnote{We are aware that the velocity profile of the wind may differ from the one we are assuming here. Specifically, the wind may be accelerated by radiation pressure or MHD driving, showing a non-linear dependence both on the radius $r$ and on the azimuthal angle with respect to the accretion disk. We will check these dependences in a forthcoming work.} Given the observed UFO absorption feature, we require that $i<\theta_{out}$, that is, that the LOS has to lie inside the wind. We also consider a more refined version of the model, allowing for an internal cavity with variable angular aperture $\theta_{in}$. 

The conical geometry of the wind is approximated  by a large number $l$ (i.e.,100) of conical shells, equally spaced between $r=0$ and $r=r_{cone}$. These shells have angular opening $\theta_{out}$ with respect to the symmetry axis. For each shell, we assign angular coordinates to a number $n$ of points ($n=10^4$ by default) through a Monte Carlo method. The radial velocity is a function of radius, according to the linear relation of Eq. \ref{eq_velocity}, in which $s$ ranges from 0 to 1 and represents the deceleration factor of the wind as a function of $r'$, which is the distance from the vertex normalized to the height of the cone ($r_{cone}$). The latter is not a parameter of the WINE model, as in Eq. \ref{eq_velocity} the radial velocity only depends on the dimensionless radius $r' \equiv r/r_{cone}$  spanning the range [0, 1].
Accordingly, the radial velocity of the wind at the vertex is $v_{max}$, while the velocity at $r=r_{cone}$ is defined as $v_{min}$.

We assume the same intrinsic brightness $B_{int}$ for each point all over the cone (see Eq. \ref{xi_r} and following discussion). Then, using special relativity, we calculate for each point of the shell the projected value of $B_{int}$  and $v(r)$ along the LOS: $B_{proj}$ and $v(r)_{proj}$, respectively.
Values of $B_{proj}$ are then grouped into bins of 3,600 km s$^{-1}$. In this way we obtain the spectrum of a single shell. The combination of each shell spectrum at all radii (i.e., from $r=0$ to $r=r_{cone}$) provides the global spectrum emitted by the wind.

The total observed emissivity of the wind is therefore $B_{tot}=\sum\limits_{l=1}^{100} \sum\limits_{n=1}^{10^4} B_{proj}$
and parametrized in \emph{xspec} through the normalization of the WINE model component (see Sect. \ref{Spectral analysis}).
Notwithstanding the fact that $B_{int}$ can be estimated from $B_{tot}$, it is not possible to derive from it a reliable value for the Fe ions' emissivity, given the current stage of the WINE model, which would allow us to infer the column density of the wind. This would require an accurate treatment of the ionic abundances and emissivities, as well as the photoionization equilibrium and radiative transfer within the medium. This will be accounted for in a future version of the WINE model by incorporating XSTAR tables into the present modelization of the wind (see Sect. \ref{Discussion}).

Our model  assumes a radial density profile of the wind $\rho \propto r^{\alpha}$. In the case of the UFO in PDS 456, we choose $\alpha=-2$ , an isothermal density profile. This ensures that the ionization parameter as a function of the radius of the wind, $\xi(r)$, is constant at all radii, in accordance with the observations of  PDS 456 showing that the wind ionization is dominated by Fe XXVI (\citealp{reeves03}, \citealp{reeves09}, N15). In fact,  the number density of the wind can be written as
\begin{equation}
n(r)=n_0 \Big( \frac{r_0}{r} \Big) ^2
\label{n_profile}
,\end{equation}
where $n_0$ is the density of the wind at a fiducial radius $r_0$. Using Eq. \ref{n_profile}, $\xi(r)$ can be expressed as
\begin{equation}
\xi=\frac {L_{ion}}{n(r) r^2}=\frac {L_{ion}}{n_0 r_0^2}
\label{xi_r}
,\end{equation}
where there is no dependence on $r$. A constant $\xi$, as shown in Eq. \ref{xi_r}, indicates that the relative abundance of the ions, including Fe XXVI, is roughly constant within the medium and therefore justifies our assumption of constant $B_{int}$.
 
 In the following we provide a detailed description of the free parameters of the WINE model:
\begin{itemize}[noitemsep]
\item $v_{max}$: the maximum velocity of the wind (in units of \textrm{c}) corresponding to the starting velocity at the vertex of the cone. We assume [0.1c, 0.6c] as the possible range of $v_{max}$ and we span this interval with a resolution of 0.01\textrm{c}.
\item The opening angle $\theta_{out}$ can range from 0 to 90 \textrm{deg} (see Fig. \ref{Figmodelsketch}). This interval is sampled by nine equally-space steps of 10 \textrm{deg}.
\item $i$: the upper limit of this parameter is set by the requirement of having an LOS inside the cone aperture, that is, $i\leq \theta_{out}$. The possible range for $i$, [0, $\theta_{out}$], is sampled with 11 equally spaced steps.
\item The deceleration factor of the wind $s$ (Eq. \ref{eq_velocity}) ranges from 0 to 1 and it is sampled through steps of 0.1.
\item Furthermore, an inner cavity in the cone can be accounted for by an additional parameter $\theta_{in}$ , which represents the angular amplitude with respect to the symmetry axis. Parameter $\theta_{in}$ can span the open interval (0, $i$).
\end{itemize}
From this set of free parameters, it is possible to derive \emph{(i)} $v(r)$, as defined by Eq. \ref{eq_velocity}, and \emph{(ii)} $C_f$, the covering factor of the wind defined as  the fraction of sky covered by the cone, as seen from the vertex:
\begin{equation}
C_f = 1-\cos(\theta_{out})
\label{Cf}
.\end{equation}
In the case of a cone with an inner cavity, Eq. \ref{Cf} becomes
\begin{equation}
C_f = \cos(\theta_{in})-\cos(\theta_{out})
.\end{equation}

\section{Spectral analysis}
\label{Spectral analysis}

We merge the simultaneous XMM-Newton+NuSTAR spectra from Epoch 3 (September 15, 2013) and Epoch 4 (September 20, 2013) in N15 (see Table \ref{table:data}) in order to increase the signal-to-noise ratio. We consider these two datasets  since they are virtually indistinguishable in terms of flux and spectral shape. 

Data reduction and spectral extraction are performed using the data analysis software SAS 16.0.0 and NUSTARDAS v1.7.1 (calibration database version 20170222) for XMM-Newton and NuSTAR, respectively. Table \ref{table:data} lists the journal of the observations. We adopt the same filtering scheme and spectral extraction regions as in N15.

We co-add the spectra of Epochs 3 and 4, both for XMM-Newton and for NuSTAR, using the \emph{ftools} task \emph{addascaspec}, and finally group them to a minimum of 100 and 50 counts per energy bin, respectively.
\begin{table*} [!ht]
\begin{center}
\centering
\begin{tabular}{c | c c c c c c c c}
\hline\hline
Epoch & Telescope & Obs. ID & Date and time & $T_{tot}$ (ks)$^{(a)}$ & Instrument & $T_{net}$ (ks)$^{(b)}$ & Net counts (k)$^{(c)}$ & Extraction$^{(d)}$ \\
\hline
3 & XMM & 0721010501 & 2013-09-15 18:30:00 & 120.5 & pn & 102.2 & 207.6 & 35/60\\
 & NuSTAR & 60002032006 & 2013-09-15 17:56:07 & 119.1 & FPMA & 44.0 & 4.3 & 80/80\\
 &  & && & FPMB & 44.0 & 4.0 & 80/80\\
 \hline
4 & XMM & 0721010501 & 2013-09-20 02:29:39 & 112.1 & pn & 94.9 & 188.5 & 35/60\\
 & NuSTAR & 60002032008 & 2013-09-20 03:06:07 &113.8 & FPMA &  58.5 & 5.9 & 80/80\\
 &  &  &  & & FPMB & 58.5 & 5.7 & 80/80\\
\hline
\end{tabular}
\end{center}
\caption{Journal of the observations. $(a),(b)$ Observation lengths, in \textrm{ks}, before and after filtering, respectively. $(c)$ Net counts in units of $10^3$photons in the 3-10(30) keV band for XMM-Newton(NuSTAR). $(d)$ Extraction regions for source and/or background in \textrm{arcsec}.}
\label{table:data}
\end{table*}
 We base our fitting procedure on that of N15, with a lower energy bound of 3 \textrm{keV}. This allows us to avoid two complex spectral features: the soft excess at energies below $\sim 1$ \textrm{keV} and the warm absorber at $\lesssim 2$ \textrm{keV}. The upper bounds are 10 and 30 \textrm{keV} for the XMM-Newton and NuSTAR spectra, respectively. The spectra are always jointly fitted, with an intercalibration constant left free to vary.

\subsection{Phenomenological fit}
 \label{fen fit}
\begin{figure}
\centering
\includegraphics[width=9cm]{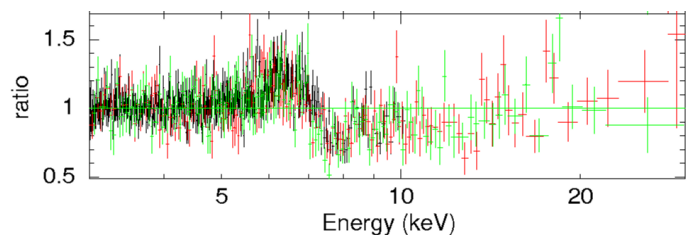}
\caption{Ratio between data and the best-fit continuum model to the 3-5 keV + 15-30 keV spectrum, once the 5-14 \keV\ region is included (see Sect. \ref{fen fit} for details). Black, green, and red crosses indicate XMM-Newton/pn, NuSTAR FPMA, and FPMB data.}
\label{FigContinuum}
\end{figure}

Similarly to N15, we first perform a preliminary fit to describe the continuum emission using a power law modified by a neutral partial covering absorber, responsible for the spectral curvature below 4 \textrm{keV}.  We ignore the energy interval around the Fe K shell features, from $5$ and $14$ \textrm{keV}. Using directly the \emph{xspec} notation, the fitting model can be expressed by the  analytical expression
\begin{equation}
constant*phabs*zpcfabs*zpowerlaw
\label{Continuum model}
,\end{equation}
where $constant$ is a constant multiplicative factor, set $\equiv 1$ for XMM-Newton data and left free to vary for the two NuSTAR modules. This is necessary to account for intercalibration differences between the three instruments. Galactic absorption is represented through $phabs$, which is set to $N_H^{Gal} \equiv 2\times 10^{21}$ \textrm{cm}$^{-2}$ (see N15). The neutral partial covering absorber is described by $zpcfabs$. Finally, the spectral index $\Gamma$ and normalization \textit{K} of the power-law continuum are left free to vary.
\begin{table} [!ht]
\begin{center}
\centering
\begin{tabular}{l c}
\hline\hline
Parameter & Value \\
\hline
\textbf{\textit{constant}}$\mathbf{^{(a)}}$ & $1^{(f)},\ 1.07\pm0.02,\ 1.08\pm0.02$\\
\textbf{\textit{phabs}} &       \\
$N_H$ & $0.2^{(f)}$ $\times 10^{22} $ \textrm{cm}$^{-2}$         \\
\textbf{\textit{zpcfabs}}&      \\
$N_H$ & $24^{+8}_{-9}$ $\times 10^{22} $ \textrm{cm}$^{-2}$\\
$f_{cov}^{\ (b)}$& $0.42^{+0.05}_{-0.04}$\\
\textbf{\textit{zpowerlaw}}&\\
$\Gamma$& $2.40\pm0.07$\\
$K^{(c)}$ & $5.4^{+1.3}_{-1.0}$\\
$z$ & $0.184^{(f)}$ \\
\hline
$\mathbf{\chi^2 /d.o.f}.$&309/362\\
\hline
\end{tabular}
\end{center}
\caption{Best-fit values for the continuum fitting (see relation \ref{Continuum model}). Hereafter, errors are at $68\%$ c.l.. $^{(a)}$ \textit{constant} has three values because is the only parameter allowed to vary between the different datasets (values shown for XMM-Newton pn and NuSTAR FPMA and FPMB, respectively). $^{(b)}$ $f_{cov}$ is the covering fraction of  the intrisic cold absorber. $^{(c)}$ Normalization in units of $10^{-3}$photons \textrm{keV}$^{-1}$ \textrm{cm}$^{-2}$\textrm{s}$^{-1}$ at 1 \textrm{keV}.  $^{(f)}$ Fixed value.}
\label{table:continuum}
\end{table}
We obtain a spectral index $\Gamma = 2.40 \pm 0.06$ for the continuum component. For the absorber, we find a column density of $N_H=2.4\pm0.3\times 10^{23} $ \textrm{cm}$^{-2}$ and covering fraction of $C_f=0.42\pm0.05$,  suggesting a more distant feature from the AGN with respect to the UFO. These findings are in agreement with the low values for $N_H$ and $\xi$ and the low variability found in N15 for this component. The resulting $\chi^2$ is 309 for 362 degrees of freedom. Table \ref{table:continuum} shows the best-fit parameters.
Figure \ref{FigContinuum} shows the residuals in terms of data-to-model ratios over the 3-30 keV band.

Then, in order to have a first description of the absorption and emission features related to the Fe K shell, we consider the whole energy band adding a Gaussian emission line with energy corresponding to Fe XXVI \lya\ ($E=6.97$ \textrm{keV} rest frame) and free width and redshift. We model the absorption including two Gaussian lines for Fe XXVI \lya\ and \lyb\ ($E=8.25$ \textrm{keV} rest frame) and a photoelectric edge at $E=9.28$ \textrm{keV} rest frame, to take into account the K shell edge from Fe XXVI. 

The inclusion of these spectral components is based on the phenomenological analysis reported in N15, in order to fully characterize the observed Fe K spectral features. The resulting model is
\begin{equation}
\begin{split}
constant*phabs*(zpcfabs*zedge*zpowerlaw\\
+zgauss\_ em\_Ly\alpha+zgauss\_abs\_Ly\alpha+zgauss\_abs\_Ly\beta)
\label{fen model}
\end{split}
.\end{equation}
In addition to the components in expression \ref{Continuum model}, $zgauss\_ em\_Ly\alpha$ indicates the Gaussian emission associated to Fe XXVI \lya\ and $zgauss\_abs\_Ly\alpha$, $zgauss\_abs\_Ly\beta$ represents the Gaussian absorption for \lya\ and \lyb\ lines, respectively. The term $zedge$ corresponds to the photoelectric edge due to the Fe K shell.

A simple fit leaving all the parameters free returns an almost identical redshift for both the absorption lines,  $z=-0.100\pm0.009$, $-0.107\pm0.005$, for \lya\ and \lyb\  , respectively, corresponding to a blueshifted velocity of $\simeq0.26, 0.27$\textrm{c}, while the redshift of the K shell edge,  $z=-0.071\pm0.1$, implies a velocity of $\simeq 0.24$\textrm{c}.
Since we expect that the same medium is responsible for all the absorption features, we impose the same redshift for the lines and the edge. Moreover, we link the Gaussian line widths in order to have the same velocity dispersion.
Table \ref{table:gauss} lists the best-fit values for these emission and absorption components, obtained by keeping the continuum and the partial covering absorption parameters fixed to the values shown in Table \ref{table:continuum}. The best-fit results in a $\chi^2=649$ for 705 degrees of freedom (d.o.f.), confirming the validity of our hypothesis.

Interestingly, the redshift of the emitting component is significantly shifted with respect of that of the host galaxy: $z=0.117$ versus $z=0.184$. This corresponds to a difference in terms of blue-shifted velocity along the LOS of $\simeq0.06$\textrm{c} ($\sim 18,000$ \textrm{km s}$^{-1}$), lending support to the idea that we are observing the approaching component of the biconical outflow as in Fig. \ref{Figmodelsketch}.

\begin{table} [!ht]
\begin{center}
\centering
\begin{tabular}{l c }
\hline\hline
Parameter & Value \\
\hline
\textbf{\textit{zedge}} &\\
$Edge\ En.$ (\textrm{keV}) &$9.28^{(f)}$\\
$f^{(a)}$ &$0.19\pm0.04$\\
$z^{(b)}$ &$-0.096^{+0.008}_{-0.007}$\\
\textbf{\textit{zgauss\_ em}}&\\
$En.$  (\textrm{keV}) & $6.97^{(f)}$\\
$\sigma$ (\textrm{keV}) & $0.7\pm 0.1$\\
$z$ & $0.12^{+0.01}_{-0.02}$\\
$K^{(c)}$ & $11^{+2}_{-1}$\\
$\mathbf{zgauss\_Ly\alpha}$&\\
$En.$ (\textrm{keV}) & $6.97^{(f)}$\\
$\sigma^{(d)}$  (\textrm{keV}) & $0.33\pm 0.04$\\
$z^{(b)}$ &$-0.0962^{+0.008}_{-0.007}$\\
$K^{(c)}$& $5.8^{+0.6}_{-0.1}$\\
$\mathbf{zgauss\_Ly\beta}$&\\
$En.$ (\textrm{keV}) & $8.25^{(f)}$\\
$\sigma^{(d)}$  (\textrm{keV}) &$0.39\pm 0.05$\\
$z^{(b)}$ &$-0.0962^{+0.008}_{-0.007}$\\
$K^{(c)}$& $2.6\pm0.4$\\
\hline
$\mathbf{\chi^2 /d.o.f}.$&649/705\\
\hline
\end{tabular}
\end{center}
\caption{Best-fit values for the phenomenological fit (see relation \ref{fen model}). $^{(a)}$ $f$ is defined as the absorption depth at threshold. $^{(b)}$ Linked values. $^{(c)}$ Gaussian function normalization $K$ is in units of $10^{-6}$photons \textrm{cm}$^{-2}$ \textrm{s}$^{-1}$. $^{(d)}$ Linked values. $^{(f)}$ Fixed value.}
\label{table:gauss}
\end{table}

\subsection{Fit with the WINE model}
\label{Results}
As a further step, we replace the Gaussian emission line of the phenomenological fit with two WINE model components, corresponding to \lya\ and \lyb\ emission. The continuum and neutral partial covering absorber are still parametrized as in Sect. \ref{fen fit}.

The fitting model is described by the  expression
\begin{equation}
\begin{split}
constant*phabs*(zpcfabs*smedge*zpowerlaw\\
+ WINE\_Ly\alpha + WINE\_Ly\beta +zgauss\_Ly\alpha+zgauss\_Ly\beta)
\label{Complete model}
\end{split}
,\end{equation}
where $WINE\_Ly\alpha$ and $WINE\_Ly\beta$ represent the two WINE model emission components. The two Gaussian absorption lines, $zgauss\_Ly\alpha$ and $zgauss\_Ly\beta$, account for \lya\ and \lyb\ absorption, respectively, and $smedge$ (\citealp{smedge}) is a smeared photoelectric edge, to account for the velocity gradient of the wind (see Eq. \ref{eq_velocity}).
$WINE\_Ly\alpha$ and $WINE\_Ly\beta$ have identical parameters, except for rest-frame energy and normalization. Specifically, the ratio between the normalizations corresponds to the ratio of the oscillator strength of Fe XXVI \lya\ and \lyb\ (\citealp{molendi}, \citealp{tombesi osc}).

Since we expect that both the emission and absorption features are due to the same medium, we tie the redshift and broadening of $zgauss\_Ly\alpha$, $zgauss\_Ly\beta,$ and $smedge$ to the emission parameters in the WINE model. Specifically, since our LOS intercepts the UFO, the velocity component of the gas along the LOS (i.e., the gas responsible for the absorption) coincides with the radial velocity $v(r)$. So, the average redshift $z_{avg}$ of $zgauss\_Ly\alpha$, $zgauss\_Ly\beta$ and $smedge$ corresponds to the average outflow velocity, $v_{avg}  \equiv v\ (r_{cone}/2)$, according to the  equation
\begin{equation}
z_{avg}=\frac {1+z_{PDS}}{\sqrt{(1+v_{avg})/(1-v_{avg})}}-1
\label{z avg}
,\end{equation}
where $z_{PDS}=0.184$. Moreover, we assume that the standard deviation $\sigma$ of $zgauss\_Ly\alpha$ and $zgauss\_Ly\beta$ is dominated by the velocity shear of the wind. Hence, we set $\sigma$, in terms of velocity, equal to the difference $v_{max}-v_{avg}$, while in terms of energy it can be expressed as
\begin{equation}
\sigma=E* \left( \frac {1}{1.184\sqrt{(1-v_{max})/(1+v_{max})}}-\frac{1}{1+z_{avg}} \right)
\label{sigma}
,\end{equation}
where $E$ is the rest frame line energy. The same formula, using the photoionization energy threshold, can be used to express the smearing characteristic width in the $smedge$ component.

\begin{figure}
\centering
\includegraphics[width=9cm]{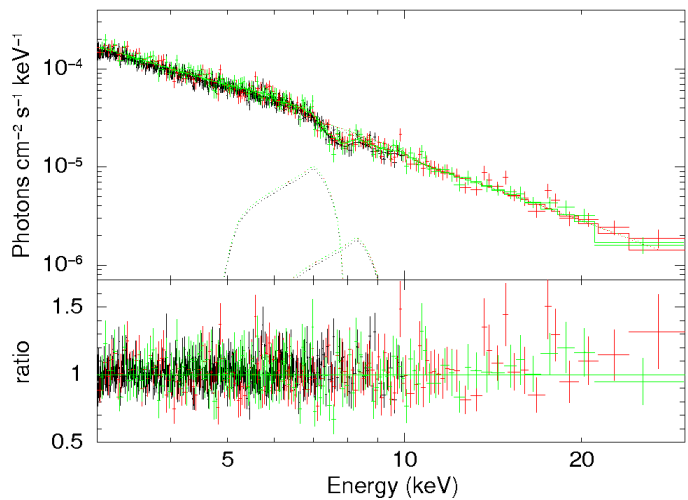}
\caption{Top panel: data (symbols as in fig \ref{FigContinuum}) and best-fit model (solid line) including two WINE model components (dotted lines). Bottom panel: ratios between data and best fit model. See Sect. \ref{Results} for more details. }
\label{Fignewcav}
\end{figure}

Figure \ref{Fignewcav} shows the result of the fit assuming a full cone (without inner cavity). The associated $\chi ^2$ (dof) is 652 (715).
The best-fit values are listed in Table \ref{table:nocav}.
The radial velocity $v(r)$ ranges from $\simeq 0.28$\textrm{c} to $\simeq 0.17$\textrm{c}. Regarding the outflow geometry, we obtain $\theta_{out}\sim 70$ deg, $i \sim 60$ deg, and a covering fraction (calculated as the fraction of visible hemisphere covered by the wind) of $C_f \sim 0.7$.

We also evaluate the inclusion of the cavity and find that it is not statistically required by the data, with a poorly constrained value of $\theta_{in}$, with an upper limit of $21$ deg (90$\%\ c.l.$).

\begin{table} [!ht]
\begin{center}
\centering
\begin{tabular}{l l }
\hline\hline
Parameter & Value\\
\hline
\textbf{\textit{smedge}} &\\
$E$ (\textrm{keV}) &$9.28^{(f)}$\\
$z^{(a)}$ & $-0.063^{+0.004}_{-0.005}$\\
$f^{(b)}$ &$0.23\pm0.05$\\
$index^{(c)}$ &$-2.67^{(f)}$\\
$width^{(d)}\ (\textrm{keV})$ &$0.60^{+0.08}_{-0.09}$\\

\textbf{\textit{WINE model}}&\\
$v_{max}$ (\textrm{c}) & $0.285^{+0.006}_{-0.007}$\\
$s$ & $0.39^{+0.03}_{-0.04}$\\
$\theta_{out}$ (\textrm{deg}) & $71^{+13}_{-8}$\\
$i$ (\textrm{deg}) & $63^{+13}_{-16}$\\
$K^{(e)(g)}$ ( \lya\ ) & $1.2^{+0.3}_{-0.2}\times 10^{-7}$\\
$K^{(e)(g)}$ ( \lyb\ ) & $2.28^{+0.6}_{-0.3}\times 10^{-8}$\\
 $z$ & $0.184^{(f)}$\\
 \hline
 $v_{min}^{(h)}$ (\textrm{c}) & $0.17\pm 0.01$\\
$C_f^{(i)}$ & $0.7^{+0.2}_{-0.1}$\\
\hline
 
$\mathbf{zgauss\_Ly\alpha}$&\\
$E$ (\textrm{keV}) & $6.97^{(f)}$\\
$\sigma^{(d)}$ (\textrm{keV}) & $0.45^{+0.06}_{-0.07}$\\
$z^{(a)}$ & $-0.063^{+0.004}_{-0.005}$\\
$K^{(e)}$& $1.28\pm 0.2\times 10^{-5}$\\

$\mathbf{zgauss\_Ly\beta}$&\\
$E$ (\textrm{keV}) & $8.25^{(f)}$\\
$\sigma^{(d)}$ (\textrm{keV}) &$0.54^{+0.07}_{-0.08}$\\
$z^{(a)}$ & $-0.063^{+0.004}_{-0.005}$\\
$K^{(e)}$& $4.0^{+0.6}_{-0.7}\times 10^{-6}$\\
\hline
$\mathbf{\chi^2/d.o.f.}$ & 652/715\\
\hline 
\end{tabular}
\end{center}
\caption{Best-fit values for the absorption and emission components using the WINE model (see relation \ref{Complete model}). $^{(a)}$ Redshift of the absorption components are linked. $^{(b)}$ $f$ is defined as the absorption depth at threshold. $^{(c)}$ Photoelectric cross-section index. $^{(d)}$ Linked values. $^{(e)}$ Normalizations in units of total photons \textrm{cm}$^{-2}$ \textrm{s}$^{-1}$. $^{(g)}$ WINE model normalizations are linked according to the ratio of the \lya\ and \lyb\ oscillator strengths. $^{(h)}$ Derived minimum velocity of the wind (i.e., $v(r)$ for $r=0$ and $r=r_{cone}$). $^{(i)}$ Derived opening angle of the cone. $^{(f)}$ Fixed value.}
\label{table:nocav}
\end{table}

\subsection{Reliability of the WINE model}
We perform extensive simulations in order to check the reliability of the WINE model. In particular here we present two cases. We first use as input our best-fit values (see Sect. \ref{Results} and Table \ref{table:nocav}) and simulate XMM-Newton and NuSTAR FPMA and FPMB spectra with the same summed (Epoch 3 + Epoch 4) net exposure times reported in Table \ref{table:data}. We adopt the same spectral grouping as described in Sect. \ref{Spectral analysis}. For each simulation, we first perform a fit in the range 3-10(3-30) \keV\  for XMM-Newton(NuSTAR), ignoring the interval between 5 and 14 \textrm{keV}, to find the best-fit continuum and cold absorber parameters. Then, we freeze these parameters and we fit the WINE model (together with the two Gaussian absorption lines and the photoelectric edge), considering also the energy interval 5-14 \textrm{keV}.

Table \ref{Table:reliability}, first column, reports the differences $\Delta$ between the input values and the results, according to the  formula
\begin{equation}
\Delta= \frac { | V_i - \overline{V}_{out} | } {\sigma_{out}}
\label{errors}
,\end{equation}
 where $V_i$ is the input value, $\overline{V}_{out}$ is the mean of the distribution of the best-fit values from 100 simulated spectra, and $\sigma_{out}$ is the standard deviation of the distribution. The agreement between the mean values and the input parameters is remarkably good.
We then perform an additional set of simulations, using different arbitrary input values, to check the capability of the WINE model to discriminate between different scenarios. Columns 2 and 3 of Table \ref{Table:reliability} report the input values and the resulting $\Delta$, respectively.
\begin{table} [!ht]
\begin{center}
\centering
\begin{tabular}{l c c c}
\hline\hline
\  & Model 1$^{(a)}$ & Model 2$^{(b)}$ \\
\ Parameter & $\Delta$ & Input & $\Delta$ \\ 
\hline
$v_{max}$ & 0.07 & 0.35 \textrm{c} & 0.05 \\
$s$ & $0.37$ & 0.7 & 0.21\\
$\theta_{out}$ & 0.27 & 45 \textrm{deg} & 0.31 \\
$i$ & 0.44 & 4.5  \textrm{deg} & 0.47\\
\hline
\ Derived qts.$^{(c)}$\\
$v_{min}$ & 0.36 & 0.10 \textrm{c} & 0.21\\
$C_f$ & 0.30 & 0.29 & 0.26\\
\hline
\end{tabular}
\end{center}
\caption{$^{(a)}$ Difference $\Delta$ between input values and results (see Eq. \ref{errors}). Input parameters are those reported in Table \ref{table:nocav}). $^{(b)}$ Different arbitrary input values (center column) and corresponding $\Delta$ (right column). $^{(c)}$ These quantities are derived from the fit parameters, as described in Sect. \ref{Emission model}.}
\label{Table:reliability}
\end{table}

\section{Discussion}
\label{Discussion}

Using the novel wind emission (WINE) model, we constrained the velocity (mean value $\sim 0.23$c), opening angle (71 deg), and covering fraction (0.7) of the UFO in the quasar PDS 456. The results of our analysis are in agreement with the estimates on the bulk wind velocity and $C_f$ in N15. For the first quantity they estimated $\sim 0.25$\textrm{c}, which is inside our radial excursion ($0.17\ -\ 0.28$\textrm{c}). 

The covering fraction in N15 is evaluated in different ways. The normalization of the XSTAR wind emission tables yields an average value between all the observations of $C_f=0.8\pm0.1$, while the fraction of the absorbed continuum luminosity re-emitted by the wind gives $C_f>0.5$. 

We find a wind opening angle of $71^{+13}_{ \  -8}$ \textrm{deg}, implying $C_f=0.7^{+0.2}_{-0.1}$. 
This value is consistent with N15 and with UFO detection rates in the larger AGN population (see, e.g., \citealp{tombesi}, \citealp{cf1}, \citealp{cf2}). 

Covering fraction is a key property in the calculation of the mass outflow rate, as expressed in \citet{crenshaw},
\begin{equation}
\dot M_{out} = 4 \pi r N_H \mu m_p C_f v
,\end{equation}
with $r$ the launching radius, $\mu$ the mean atomic mass per proton ($\approx1.2$, \citealp{mu}), $m_p$ the proton mass, and $v$ the outflow velocity. From $\dot M_{out}$ and $v$, it is possible to estimate the momentum rate, $\dot P_{out} = \dot M_{out} v$, and the energy transfer rate, $\dot E_{out}=\frac {1}{2}\dot M_{out} v^2$. These quantities are fundamental in the determination of the AGN feedback towards the host galaxy, and hence of their coupled evolution.

For the outflow velocity, we use $v=(0.23\pm 0.06)$c, that is, the average velocity of the UFO $v_{avg}$ as the mean value, and $v_{max}$ and $v_{min}$ (i.e., $v(r_{cone})$) as upper and lower limit, respectively.
Moreover, we consider our covering fraction $C_f=0.7$, while the following quantities are taken from N15: $N_H=6\times 10^{23} $ \textrm{cm}$^{-2}$, $r=100 r_g\ (=1.5\times 10^{16}$ \textrm{cm}), $L_{bol}\sim 10^{47}$ \textrm{erg s$^{-1}$} , and the black hole mass $M_{BH}=10^9 M_{\odot}$.
We find $\dot M_{out}\sim 16\pm 4$ \textrm{M$_ {sun}$ yr$^{-1}$}$\sim 0.23\pm 0.06 \dot M_{Edd}$, for a radiative efficiency $\eta =0.3$, as expected for luminous quasars such as PDS 456 (see, e.g., \citealp{eta1}, \citealp{eta2}). We derive a momentum rate $\dot P_{out} = 7\pm 3\times 10^{36}$ \textrm{dyne}, $\sim 2.1\pm 1.1$ times the radiation momentum rate $\dot P_{rad}=L_{bol}/$\textrm{c}, and an energy rate $\dot E_{out}=3\pm 2 \times 10^{46}$ \textrm{erg s$^{-1}$}, that is, $\sim\ 30\pm 20 \%\ L_{AGN}$. 

The average values for $\dot M_{out}$ and $\dot E_{out}$ are a factor of $\sim 1.5$ and $\sim 1.3$ times greater than that found in N15, respectively. 
This is mainly due to the different covering fractions (0.7 versus 0.5) and because they assume $\mu \equiv 1$ (i.e., the gas is composed only by hydrogen). 

To investigate the possibility that the quality of the observations did not allow us to constrain the presence of an inner cavity, we run a set of 100 simulations of observation with the X-ray Integral Field Unit (X-IFU) instrument of the Advanced Telescope for High-Energy Astrophysics (ATHENA) \citep{nandra}.
We adopt the best fit model (see \ref{Complete model} and Table \ref{table:nocav}) , including an internal cavity with angular amplitude $\theta_{in}=21$ \textrm{deg}, the $90\%$ upper limit reported in Sect. \ref{Results}.  We find that 500 \textrm{ks} are necessary to constrain the presence of the cavity, with a confidence level of 90$\%$. With this exposure time we can measure the parameters with an accuracy higher than 6$\%$, except for $\theta_{in}$ and $C_f$, for which the relative uncertainties are $18\%$ and $11\%$, respectively. Further details are in Appendix \ref{ATHENA}.

The results of this work show the robustness of the WINE model and its utility to constrain the properties of the outflow. This could be useful especially for those cases in which it is not clear whether the emission and/or absorption features are mainly due to disk reflection or nuclear winds (see, e.g., \citealp{kouichi16} for 1H 0707-495 and \citealp{FERO}, \citealp{SUZAKU} for type I AGNs). More generally, this model represents a physically and geometrically based approach to explore outflow kinematics and can shed new light also on known UFOs from quasar sources (see, e.g.,  \citealp{kouichi17}, \citealp{parker}, \citealp{IRAS}). With some minor changes, this model can be applied also to larger scale outflows, such as broad line regions (BLRs) (\citealp{WISSH IV}) up to galactic-scale outflows (\citealp{Mrk231}).

In the forthcoming version of the WINE model we will use the XSTAR code to calculate the ionic abundances and emissivities. This will allow us to accurately take into account all the relevant transitions as a function of the density and the ionization parameter of the wind. We will also be able to constrain the launching radius and the spatial extent of the wind. Accordingly, the next version of the model will self-consistently represent both the emission and absorption features.
\\

\emph{Acknowledgements.} We thank the referee for useful comments and suggestions that helped improve the quality and the presentation of the paper. EP and LZ acknowledge financial support from the Italian Space Agency (ASI) under the contract ASI-INAF I/037/12/0 (NARO): “The unprecedented NuSTAR look at AGN through broadband X-ray spectroscopy”. FT acknowledges support by the Programma per Giovani Ricercatori - anno 2014 ``Rita Levi Montalcini''. We thank Dr. K. Fukumura for helpful discussions.

\begin{appendix}

\section{ATHENA simulations}
\label{ATHENA}
In order to assess the capability of the next-generation X-ray observatory ATHENA \citep{nandra} to constrain the presence of an internal cavity, we run a series of simulations with the X-IFU instrument\footnote{We use response files and background spectra available at \url{http://x-ifu.irap.omp.eu/resources-for-users-and-x-ifu-consortium-members/}. We simulate the case of a mirror module radius of $1469\ mm$ and adopt a background spectrum for an extraction area of 5 \textrm{arcsec} radius.}, using variable exposure times. We adopt the best-fit model found in Sect. \ref{Results}, including also a cavity with $\theta_{in}=21\ deg$, the 90$\%$ upper limit estimated in Sect. \ref{Results}. We simulate a set of 100 spectra. For each one, we group the spectrum to a minimum of 100 counts per bin; then, we perform two different spectral fittings of the data, in the energy range 3-12 \textrm{keV}. In the first one, we assume a full cone geometry and in the second one we include the inner cavity. Finally, we perform the F-test to check if the difference in the statistics of the two fits is enough to justify the introduction of the cavity.
We verify that an exposure time of 500 $ks$ ensures that 90$\%$ of the simulated spectra allows us to discriminate the presence of the cavity with an F-test probability $<1\%$. 

\begin{figure}
\centering
\includegraphics[width=9cm]{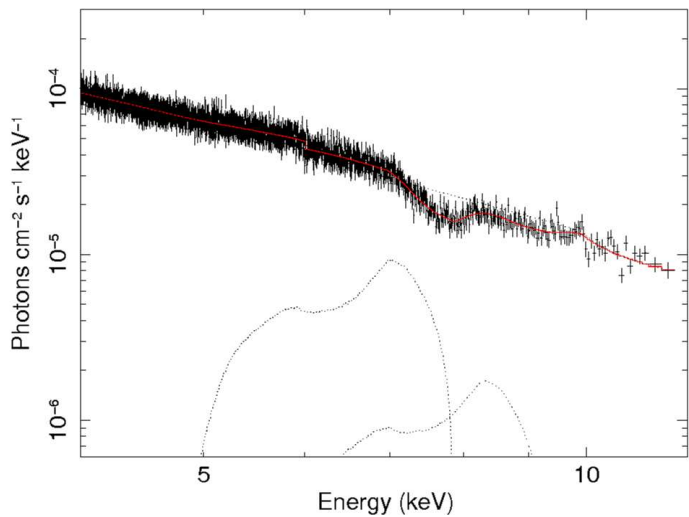}
\caption{Particular of one of the ATHENA X-IFU 500 \textrm{ks} simulated spectrum, along with the best-fit model including an inner cavity (solid red line). The two \lya\ and \lyb\ WINE model components are reported as dotted lines.}
\label{athenaplot}
\end{figure}

Table \ref{Table:athena} reports the mean and the standard deviation of the distribution of the best-fit values using an exposure of 500 $ks$. Figure \ref{athenaplot} reports one of the simulated 500 $ks$ spectra along with the best-fit inner cavity model.
\begin{table} [!ht]
\begin{center}
\centering
\begin{tabular}{l c c}
\hline\hline
\ Parameter & Input$^{(a)}$ & Results$^{(b)}$\\ 
\hline
$v_{max}\ (c)$ & $0.285$ & $0.285\pm 0.003$ \\
$s$ & $0.39$ & $0.38\pm 0.01$ \\
$\theta_{out}\ (deg)$ & $71$ & $66\pm 4$ \\
$i\ (deg)$ & $63$ & $62\pm 3$\\
$\theta_{in}\ (deg)$ & $21$ & $22\pm 4$ \\
\hline
Derived qts.$^{(c)}$\\
$v_{min}\ (c)$ & $0.174$ & $0.176\pm 0.005$ \\
$C_f$ & $0.60$ & $0.53\pm 0.06$ \\
\hline
\end{tabular}
\end{center}
\caption{$^{(a)}$ Input values. $^{(b)}$ Mean $\pm$ standard deviation for the distribution of the best-fit values for the 100 simulated spectra, with an observation time of 500 $ks$. $^{(c)}$ These quantities are derived from the fit parameters, as described in Sect. \ref{Emission model}.}
\label{Table:athena}
\end{table}

\end{appendix}


\begin{thebibliography}{}

\bibitem[Bischetti et al.(2017)]{WISSH I} Bischetti et al., 2017, A\&A, 598, A122

\bibitem[Blandford \& Payne(1982)]{blandford} Blandford \& Payne, 1982, MNRAS, 199, 883

\bibitem[de La Calle Pérez et al.(2010)]{FERO} de La Calle Pérez et al., 2010, A\&A, 524, A50

\bibitem[Crenshaw \& Kraemer(2012)]{crenshaw} Crenshaw \& Kraemer, 2012, ApJ, 753, 1, 75

\bibitem[Davis \& Laor(2011)]{eta1} Davis \& Laor, 2011, ApJ, 728, 98

\bibitem[Di Matteo et al.(2005)]{di matteo} Di Matteo et al., 2005, Nature 433, 604

\bibitem[Ebisawa et al.(1994)]{smedge} Ebisawa et al.\ 1994, \pasj, 46, 375 

\bibitem[Elvis(2000)]{elvis00} Elvis, 2000, ApJ, 545, 63-76

\bibitem[Faucher-Giguère \& Quataert(2012)]{faucher} Faucher-Giguère \& Quataert, 2012, MNRAS, 425, 605

\bibitem[Feruglio et al.(2015)]{Mrk231} Feruglio et al., 2015, A\&A. 583, A99

\bibitem[Fiore et al.(2017)]{ff17} Fiore et al., 2017, A\&A, 601, A143

\bibitem[Fukumura et al.(2010)]{fukumura10} Fukumura et al., 2010, ApJ, 715, 1

\bibitem[Fukumura et al.(2014)]{fukumura} Fukumura et al., 2014, ApJ, 780, 2, 120  

\bibitem[Gofford et al.(2013)]{cf1} Gofford et al., 2013, MNRAS, 430, 1, 60-80

\bibitem[Gofford et al.(2015)]{mu} Gofford et al.\ 2015, \mnras, 451, 4169 

\bibitem[Kallman \& Bautista(2001)]{kallman} Kallman \& Bautista, 2001, ApJS, 133, 221  

\bibitem[Kazanas et al.(2012)]{kazanas} Kazanas et al., 2012, Astron. Rev., 7, 3

\bibitem[King \& Pounds(2015)]{king_pounds15} King \& Pounds, 2015, Ann. Rev. Astron. and Astrophys, 53, 115-154

\bibitem[Kormendy \& Ho(2013)]{kormendy} Kormendy \& Ho, 2013, ARAA, 51, 1, 511-653

\bibitem[Kouichi et al.(2016)]{kouichi16} Kouichi et al., 2016, MNRAS, 461, 4

\bibitem[Kouichi et al.(2017)]{kouichi17} Kouichi et al., 2017, MNRAS, 468, 2

\bibitem[Molendi et al.(2003)]{molendi} Molendi et al., \ 2003, \mnras, 343, L1 

\bibitem[Nandra et al.(2013)]{nandra} Nandra et al., 2013, arXiv:1306.2307

\bibitem[Nardini et al.(2015)]{nardini} Nardini et al., 2015, Science, 347,6224, 860-863    

\bibitem[Oshuga et al.(2009)]{oshuga09} Oshuga et al., 2009, PASJ, 61, 3, L7-L11

\bibitem[Parker et al.(2017)]{parker} Parker et al., 2017, NATURE, 543, 7643

\bibitem[Patrick et al.(2012)]{SUZAKU} Patrick et al., 2012, MNRAS, 426,3

\bibitem[Proga \& Kallman(2004)]{proga} Proga \& Kallman, 2004, ApJ, 616, 2, 688-695   

\bibitem[Reeves et al.(2003)]{reeves03} Reeves et al., 2003, ApJ, 593, L65

\bibitem[Reeves et al.(2009)]{reeves09} Reeves et al., 2009, ApJ, 701, 1

\bibitem[Tombesi et al.(2010)]{tombesi} Tombesi et al., 2010, A\&A, 521, A57  

\bibitem[Tombesi et al.(2011)]{tombesi osc} Tombesi et al., 2011, ApJ, 742, 1

\bibitem[Tombesi et al.(2014)]{cf2} Tombesi et al., 2014, MNRAS, 443, 3, 2154-2182

\bibitem[Tombesi et al.(2015)]{tombesi15} Tombesi et al., 2015, Nature, 519, 7544

\bibitem[Tombesi et al.(2017)]{IRAS} Tombesi et al., 2017, ApJ, 850, 2

\bibitem[Trakhtenbrot(2014)]{eta2} Trakhtenbrot, 2014, ApJL, 789, 1, L9

\bibitem[Vietri et al.(2018)]{WISSH IV} Vietri et al., 2018, A\&A accepted, \url{https://arxiv.org/abs/1802.03423}

\bibitem[Zubovas \& King(2012)]{zubovas} Zubovas \& King, 2012, ApjL, 745, L34

\end{thebibliography}
\end{document}